\documentclass{elsart}   % For Latex2e
\usepackage{amsmath,amssymb}   % For Latex2e
\usepackage{graphicx}
\usepackage[all]{xy}

\begin{document}
\begin{frontmatter}

\title{Computing Spectral Sequences}
\thanks[now]{Partially supported by SEUI-MEC, project TIC2002-01626.}
\author[UR]{A. Romero\thanksref{now}},
\ead{ana.romero@dmc.unirioja.es}
\author[UR]{J. Rubio\thanksref{now}},
\ead{julio.rubio@dmc.unirioja.es}
\author[UJF]{F. Sergeraert}
\address[UR]{Departamento de Matem\'aticas y Computaci\'on.\\
 Universidad de La Rioja, Spain.  }
\address[UJF]{Institut Fourier. \\
Universit\'e Joseph Fourier. Grenoble, France}
 \ead{francis.sergeraert@ujf-grenoble.fr}

\begin{abstract}
In this paper, a set of programs enhancing the Kenzo system is presented. Kenzo is a Common Lisp program designed for computing in Algebraic Topology, in particular it allows the user to calculate homology and homotopy groups of complicated spaces. The new programs presented here entirely compute Serre and Eilenberg-Moore spectral sequences, in particular the \(E^r_{p,q}\) and \(d^r_{p,q}\) for arbitrary \(r\). They also determine when \(E^r_{p,q} = E^\infty_{p,q}\) and describe the filtration of the target homology groups \(H_{p+q}\) by the \(E^\infty_{p,q}\)'s.
\end{abstract}


\end{frontmatter}

\section{Introduction}
The computation of homology groups of topological spaces is one of the first problems in Algebraic Topology, and these groups can be difficult to reach, for example when loop spaces or classifying spaces are involved. The Kenzo program \cite{Kenzo}, a Symbolic Computation system devoted to Algebraic Topology, allows the computation of homology groups of a large class of spaces, namely the members of the class \(\mathcal{SS}_{EH}\), the class of the \emph{simplicial sets with effective homology}; for example this class is closed under the constructors \emph{loop spaces}, \emph{fibrations} and \emph{classifying spaces}. These results are obtained through the notion of \emph{object with effective homology}, using sophisticated connections between chain complexes generally \emph{not} of finite type and other chain complexes on the contrary of finite type \cite{RS02}.

Spectral Sequences are a useful tool in Algebraic Topology providing information on homology groups by successive approximations from the homology of appropriate associated complexes. A Spectral Sequence is a family of ``pages'' $\{E^r_{p,q},d^r\}$ of differential bigraded modules, each page being made of the homology groups of the preceding one. As expressed by John McCleary in \cite{Mac85}, ``\textit{knowledge of $E^r_{*,*}$ and $d^r$ determines $E^{r+1}_{*,*}$ but not $d^{r + 1}$. If we think of a spectral sequence as a black box, then the input is a differential bigraded module, usually $E^1_{*,*}$ , and, with each turn of the handle, the machine computes a successive homology according to a sequence of differentials. If some differential is unknown, then some other (any other) principle is needed to proceed.}'' In most cases, it is in fact a matter of computability: the higher differentials of the spectral sequence are mathematically defined, but their definition is not constructive, i.e., the differentials are not computable with the usually provided information.

In the case of spectral sequences associated to filtered complexes, a formal expression for the different groups $E^r_{p,q}$ (as quotients of some subgroups of the filtered complex) is known \cite[p. 327]{Mac94}, but this expression is not sufficient to compute the \(E^r_{p,q}\) when the initial filtered complex is not of finite type: if so, this initial complex cannot be installed on a machine, a frequent situation. It was proved in \cite{RS02} that the so-called \emph{effective homology} methods on the contrary give \emph{actual} algorithms computing homology groups related to the most common spectral sequences, Serre and Eilenberg-Moore, even when the initial filtered complex is not of finite type. In this paper, the process is somewhat \emph{reversed}: we use the effective homology methods to compute, as a by-product, the relevant spectral sequence, that is, the whole set of its components. The structure of this spectral sequence can give useful informations about the involved construction, for example about the present transgressions; sometimes this information is more interesting than the final homology groups.

This paper is organized as follows. In Section \ref{sec:prelim}, some preliminary concepts are introduced: specifically, in Subsection \ref{subsec:topol} we recall some basic definitions and results of Algebraic Topology, and in Subsection \ref{subsec:Kenzo} some indications about the program Kenzo are given. Section \ref{sec:programs} contains a description of the main features of the new programs (technical details, from both programming and mathematical sides, are skipped), which can be better understood by means of the examples introduced in Sections \ref{sec:examples} and \ref{sec:example2}. The paper ends with a section of conclusions and further work.  

\section{Preliminaries}
\label{sec:prelim}

\subsection{Some concepts in Algebraic Topology}
\label{subsec:topol}

The following definitions and results about some basic notions of Algebraic Topology can be found, for instance, in \cite{Mac94}.

\begin{defn}
A \emph{chain complex} is a pair $(C,d)$ where $C=\{C_n\}_{n \in \Zset}$ is a graded Abelian group and $d=\{d_n:C_n \rightarrow C_{n-1}\}_{n \in \Zset}$ (\emph{the differential map}) is a graded group homomorphism of degree -1 such that $d_{n-1} d_n= 0 \ \forall n\in \Zset$. The graded \emph{homology group} of the chain complex $C$ is $H(C)=\{H_n(C)\}_{n\in\Nset}$, with $H_n(C)=\mathrm{Ker}\ d_n / \mathrm{Im}\ d_{n+1}$. A \emph{chain complex homomorphism} $f : (A,d_A) \rightarrow (B,d_B)$ between two chain complexes $(A,d_A)$ and $(B,d_B)$ is a graded group homomorphism (degree 0) such that $ f d_A = d_B f$.
\end{defn}

\begin{note}
From now on in this paper, the chain complexes we work with are supposed to be \(\Zset \)-free, i.e. for each \(n \in \Zset\), \(C_n\) is
   a free \(\Zset\)-module.
\end{note}

\begin{defn}
A \emph{filtration} $F$ of a chain complex $(C,d)$ is a family of sub-chain complexes \(F_p C \subset C\) (that is, \(d(F_p C)\subset F_pC\)) such that 
\[
\cdots \subset F_{p-1}C_n \subset F_pC_n \subset F_{p+1}C_n \subset \cdots \quad \forall n\in \Zset 
\]
\end{defn}

\begin{note}
\label{note:flt-hom}
A filtration $F$ on $C$ induces a filtration on the graded homology group $H(C)$; let \(i_p: F_p C \hookrightarrow C\) the \(p\)-injection; then \(F_p(H(C)) = H(i_p) (H(F_p(C)))\).
\end{note}

\begin{defn}
A filtration $F$ of a chain complex $C$ is said to be \emph{bounded} if for each degree $n$ there are integers $s=s(n)<t=t(n)$ such that $F_sC_n=0$ and $F_tC_n=C_n$.
\end{defn}

\begin{defn}
A \emph{$\Zset$-bigraded module} is a family of $\Zset$-modules $E=\{E_{p,q}\}_{p,q\in \Zset}$. A \emph{differential} $d:E\rightarrow E$ of bidegree $(-r, r-1)$ is a family of homomorphisms of $\Zset$-modules $d_{p,q}:E_{p,q} \rightarrow E_{p-r, q+r-1}$ for each $p,q \in \Zset$, with $d_{p,q}\circ d_{p+r,q-r+1}=0$. The \emph{homology} of E under this differential is the bigraded module $H(E)\equiv H(E,d)=\{H_{p,q}(E)\}_{p,q\in \Zset}$ with $H_{p,q}(E)=\mathrm{Ker}\ d_{p,q} / \  \mathrm{Im}\ d_{p+r,q-r+1}$
\end{defn}

\begin{defn}
A \emph{Spectral Sequence} $E=\{E^r,d^r\}$ is a family of $\Zset$-bigraded modules $E^1$, $E^2$,\ldots, each provided with a differential $d^r=\{d^r_{p,q}\}$ of bidegree $(-r,r-1)$ and with isomorphisms $H(E^r, d^r)\cong E^{r+1}$, $r=1,2,\ldots$. 
\end{defn}

\begin{rem}
We must emphasize here that each $E^{r+1}$ in the spectral sequence is (up to isomorphism) the bigraded homology module of the preceding $(E^r,d^r)$. Therefore if we know the stage $r$ in the spectral sequence $(E^r,d^r)$ we can build the bigraded module at the stage $r+1$, $E^{r+1}$, but this cannot define the next differential \(d^{r+1}\) which therefore must be independently defined too.
\end{rem}

\begin{note}
A spectral sequence can be presented as a tower
\[
0=B^1 \subset B^2 \subset B^3 \subset \cdots \subset C^3 \subset C^2 \subset C^1=E^1
\]
of bigraded submodules of $E^1$, where $E^{r+1}=C^r / B^r$ and the differential $d^{r+1}$ can be taken as a mapping $C^r/B^r\rightarrow C^r/B^r$, with kernel $C^{r+1}/B^r$ and image $B^{r+1}/B^r$. 

We say that the module $C^{r-1}$ is the set of elements that \emph{live till stage $r$}, while $B^{r-1}$ is the module of elements that \emph{bound by stage $r$}. Let $C^\infty=\bigcap_r C^r$ the submodule of $E^1$ of elements that \emph{survive forever} and $B^\infty=\bigcap_r B^r$ the submodule of those elements which \emph{eventually bound}. It is clear that $B^\infty \subset C^\infty$ and therefore the spectral sequence determines a bigraded module:
\[
E^\infty_{p,q}=C^\infty_{p,q} / B^\infty_{p,q}
\]
which is the bigraded module that remains after the computation of the infinite sequence of successive homologies. 
\end{note}

\begin{defn}
A spectral sequence $(E^r,d^r)$ is said to \emph{converge} to a graded module $H$ (denoted by $E^1\Rightarrow H$) if there is a filtration $F$ of $H$ and for each $p$ isomorphisms $E^\infty_p \cong F_pH /F_{p-1}H$ of graded modules.
\end{defn}

\begin{thm}
\label{thm:spct-sqn}
\emph{(Theorem 3.1, Chapter XI, in \cite [p. 327]{Mac94})} Each filtration $F$ of a chain complex $(C,d)$ determines a spectral sequence $(E^r, d^r)$, defined by
\[
E^r_{p,q}=\frac{Z^r_{p,q}\cup F_{p-1}C_{p+q}}{dZ^{r-1}_{p+r-1,q-r+2}\cup F_{p-1}C_{p+q}}
\]
where $Z^r_{p,q}$ is the submodule $\left[a |\ a\in F_pC_{p+q}, d(a)\in F_{p-r}C_{p+q-1}\right]$, and $d^r:E^r_{p,q}\rightarrow E^r_{p-r,q+r-1}$ is the homomorphism induced on these subquotients by the differential map $d:C\rightarrow C$.\\
If F is bounded, $E^1\Rightarrow H(C)$; more explicitly, $E^\infty_{p,q} \cong F_p(H_{p+q}C)/F_{p-1}(H_{p+q}C)$ (with $F_p(HC)$ induced by the filtration $F$, as explained in Note \ref{note:flt-hom}).
\end{thm}

The definitions that follow include the notion of object with effective homology, which has an important role in Kenzo (for the computation of homology groups) and in our new programs. See \cite{RS02} for details.
  
\begin{defn}
\label{def:red}
A \emph{reduction} $\rho\equiv(D \Rightarrow C)$ between two
chain complexes is a triple $(f,g,h)$ where: (a) The components
$f$ and $g$ are chain complex morphisms $f: D \rightarrow C$ and $g: C \rightarrow D$; (b)
The component $h$ is a homotopy operator $h:D\rightarrow D$ (a graded group homomorphism of degree +1); (c) The following relations are satisfied:
  (1) $f  g = \mbox{id}_C$; (2)
  $g f + d_D h + h  d_D
        = \mbox{id}_D$;
  (3)~\ {$f  h = 0$;} (4) $h   g = 0$; (5) $h   h = 0$.
  
\end{defn}

\newpage
\begin{rem}
These relations express that $D$ is the direct sum of $C$ and a contractible (acyclic) complex. This decomposition is simply $D=\mathrm{Ker}\  f \oplus \mathrm{Im}\  g$, with $\mathrm{Im}\  g\cong C$ and $H(\mathrm{Ker}\  f )=0$. In particular, this implies that the graded homology groups \(H(D)\) and \(H(C)\) are canonically isomorphic. 
\end{rem}

\begin{defn}
A \emph{(strong chain) equivalence} between the complexes $C$ and $E$ (denoted by $C \Longleftrightarrow E$) is a triple $(D,\rho,\rho ')$ where $D$ is a chain complex, $\rho$ and $\rho'$ are reductions from $D$ over $C$ and  $E$ respectively:
\[
C \stackrel{\rho}{\Longleftarrow} D \stackrel{\rho'}{\Longrightarrow} E.
\]
\end{defn}

\begin{note}
An effective chain complex is essentially a free chain complex $C$ where each group $C_n$ is finitely generated, and there is an algorithm that returns a $\Zset$-base in each degree $n$ (for details, see \cite{RS02}). 
\end{note}

\begin{defn}
An \emph{object with effective homology} is a triple $(X,HC,\varepsilon)$ where $HC$ is an effective chain complex and $\varepsilon$ is a equivalence between a free chain complex canonically associated to $X$ and $HC$.
\end{defn}

\begin{note}
It is important to understand that in general the \(HC\) component of an object with effective homology is \emph{not} made of the homology groups of \(X\); this component \(HC\) is a free \(\Zset\)-chain complex of finite type, in general with a non-null  differential, allowing to \emph{compute} the homology groups of \(X\); the justification is the equivalence \(\varepsilon\).
\end{note}  

We end this section with a theorem that combines both spectral sequence and effective homology concepts. This is the main result on which the new programs are based. The proof is straightforward and is not presented here.

\begin{thm}
\label{thm:spct-eff}
Let $C$ be a filtered chain complex with effective homology $(HC,\varepsilon)$, with $\varepsilon=(D,\rho,\rho ')$, $\rho=(f,g,h)$, and $\rho '=(f',g',h')$. Let us suppose that filtrations are also defined in the chain complexes $HC$ and $D$. If the maps $f$, $f'$, $g$, and $g'$ are morphisms of filtered complexes (i.e., they are compatible with the filtrations) and both homotopies $h$ and $h'$ have order $\leq t$ (i.e. $h(F_pD),h'(F_pD)\subset F_{p+t}D \quad \forall p\in \Zset$), then the spectral sequences of the complexes $C$ and $HC$ are isomorphic for $r>t$:
\[
E(C)^r_{p,q}\cong E(HC)^r_{p,q} \quad \forall r>t
\]
\end{thm}

\subsection{The Kenzo program}
\label{subsec:Kenzo}

The Kenzo program \cite{Kenzo}, developed by the third author of this paper and some coworkers, is a Lisp 16,000 lines program devoted to Symbolic Computation in Algebraic Topology. It works with rich and complex algebraic structures (chain complexes, differential graded algebras, simplicial sets, simplicial groups, morphisms between these objects, reductions, etc.) and has obtained some results (for example homology groups of iterated loop spaces of a loop space modified by a cell attachment, components of complex Postnikov towers, etc.) which had never been determined before. 

To roughly explain the general style of Kenzo computations, let us firstly consider a didactical example. The definitions and results about Eilenberg-Mac Lane spaces \(K(\pi, n)\) that appear in this subsection can be found in \cite{May67}. The ``minimal'' simplicial model of the Eilenberg-Mac Lane space \(K(\Zset, 1)\) is defined by \(K(\Zset, 1)_n = Z^1(\Delta^n, \Zset) = \Zset^n\); an infinite number of simplices is required in every dimension \(\geq 1\). This does not prevent such an object from being installed and handled by the Kenzo program.
\begin{verbatim}
> (setf kz1 (k-z 1))
[K1 Abelian-Simplicial-Group]
\end{verbatim}

The \texttt{k-z} Kenzo function constructs the standard simplicial Eilenberg-Mac Lane space and this object is assigned to the symbol \texttt{kz1}. In ordinary mathematics notation, a 3-simplex of \texttt{kz1} could be for example \([3 | 5 | -5]\), denoted by \texttt(3 5 -5) in Kenzo. The faces of this simplex can be computed:
\begin{verbatim}
> (dotimes (i 4)
    (print (face kz1 i 3 '(3 5 -5))))
<AbSm - (5 -5)>
<AbSm - (8 -5)>
<AbSm 1 (3)>
<AbSm - (3 5)>
NIL
\end{verbatim}

You recognize the bar construction faces; in particular the face of index 2 is degenerated: \(\partial_2 [3 | -5 | 5] = \eta_1 [3]\). \emph{``Local''} (in fact simplex-wise) computations are so possible, we say this object is \emph{locally effective}. But no global information is available. For example if we try to obtain the list of non-degenerate simplices in dimension 3:
\begin{verbatim}
> (basis kz1 3)
Error: The object [K1 Abelian-Simplicial-Group] is 
  locally-effective.
\end{verbatim}

This basis in fact is \(\Zset^3\), an infinite set whose element \emph{list} cannot be explicitly stored nor displayed! So that the homology groups of \texttt{kz1} cannot be elementarily computed. But it is well known \(K(\Zset, 1)\) has the homotopy type of the circle \(S^1\); the Kenzo program knows this fact, reachable as follows. We can ask for the \underline{ef}fective \underline{h}o\underline{m}ology of \(K(\Zset,1)\):
\newpage
\begin{verbatim}
> (efhm kz1)
[K22 Homotopy-Equivalence K1 <= K1 => K16]
\end{verbatim}

A reduction \(K_1 = K(\Zset, 1) \Rightarrow K_{16}\) is constructed by Kenzo. What is \(K_{16}\)?
\begin{verbatim}
> (orgn (k 16))
(CIRCLE)
\end{verbatim}

What about the basis of this circle in dimensions 0, 1 and 2?
\begin{verbatim}
>(dotimes (i 3)
   (print (basis (k 16) i)))
(*)
(S1)
NIL
NIL
\end{verbatim}

\(\texttt{NIL} = \emptyset\) and the second \texttt{NIL} is ``technical'' (independently produced by the iterative \texttt{dotimes}). The basis are available, the boundary operators too:
\begin{verbatim}
> (? (k 16) 1 'S1)
------------------------------------------------------{CMBN 0}
--------------------------------------------------------------
\end{verbatim}

The boundary of the unique non-degenerate 1-simplex is the null combination of degree 0. So that the homology groups of \(K(\Zset, 1)\) are computable through the \emph{effective} equivalent object \(K_{16}\):
\begin{verbatim}
> (homology kz1 0 3)
Homology in dimension 0 :
Component Z
---done---

Homology in dimension 1 :
Component Z
---done---

Homology in dimension 2 :
---done---
\end{verbatim}    
    
This mechanism for computing homology groups of a chain complex through its effective homology has also been used in our new programs for the computation of spectral sequences, as explained in the next section.

\section{New programs}
\label{sec:programs}

The programs we have developed (with about 1800 lines) allow computations of spectral sequences of filtered complexes, when the \emph{effective} homology of this complex is available. The programs determine not only the groups, but also the differential maps $d^r$ in the spectral sequence, as well as the stage $r$ on which the convergence has been reached. In this section we explain the essential part of these programs, describing the functions with the same format as in the Kenzo documentation \cite{Kenzo}.  

For the development of the new module, we have only dealt with filtered chain complexes satisfying some basic properties: first, we work with filtrations that are bounded below, i.e. for each degree $n$ there is an integer $s=s(n)$ such that $F_sC_n=0$. And second, we suppose that for each $x\in C$ it is possible to define its filtration index $p=\mathrm{min}\{t\in\Zset |\ x \in F_tC\}$ (which implies that the filtration is convergent above, that is, $C=\bigcup_pF_pC$). 

The first step has been to increase the class system of Kenzo with the class \texttt{Filtered-Complex}, whose definition is:
\begin{verbatim}
 (DEFCLASS FILTERED-COMPLEX (chain-complex)
     ((flin :type Chcm-FltrIndex :initarg :flin :reader flin1)))
\end{verbatim}

This class inherits from the class \texttt{Chain-Complex}, and has one slot of its own:\par
{\leftskip=0.2cm \hangafter=1 \hangindent=0.8cm \texttt{flin} (FiLtration INdex function) a Lisp function that, from a degree $n$ and a generator $g\in C_n$, determines the filtration index $p=\mathrm{min}\{t\in\Zset |\ g \in F_tC_n\}$).\par}

We have designed this class with several functions that allow us to build filtered complexes and to obtain some useful information about them (when they are finitely generated in each degree). The description of some of these methods is showed here:\par
{\leftskip=0.3cm \hangafter=1 \hangindent=1.1cm \texttt{build-FltrChcm :cmpr} \textit{cmpr} \texttt{:basis} \textit{basis} \texttt{:bsgn} \textit{bsgn} 
\texttt{:intr-dffr}\\ \textit{intr-dffr} \texttt{:dffr-strt} \textit{dffr-strt} \texttt{:flin} \textit{flin} \texttt{:orgn} \textit{orgn}\\
The returned value is an instance of type \texttt{FILTERED-COMPLEX}. The keyword arguments are similar to those of the function \texttt{build-chcm} (that constructs a chain complex), with the new argument \texttt{flin} which is the filtration index function.\par}
\par
{\leftskip=0.3cm \hangafter=1 \hangindent=1.1cm \texttt{change-chcm-to-FltrChcm} \textit{chcm}  \texttt{:flin} \textit{flin} \texttt{:orgn} \textit{orgn}\\
Build a \texttt{FILTERED-COMPLEX} instance from an already created chain complex \textit{chcm}. The user must introduce the filtration index function and a list explaining the \emph{origin} of the object (see \cite{Kenzo} for more details about \texttt{orgn}).
\par}
\par
{\leftskip=0.3cm \hangafter=1 \hangindent=1.1cm \texttt{fltrd-basis} \textit{fltrcm degr fltr-index}\\
Return the elements of the basis of $F_pC_n$, with $C$ the \emph{effective} filtered chain complex \textit{fltrcm}, $n=$\textit{degr} and $p=$\textit{fltr-index}.\par}
\par
{\leftskip=0.3cm \hangafter=1 \hangindent=1.1cm \texttt{fltr-chcm-dffr-mtrx} \textit{fltrcm degr fltr-index}\\
Matrix of the differential application for degree \textit{degr} of the subcomplex $F_pC$, where $p=$\textit{fltr-index} and $C=$\textit{fltrcm} is an effective chain complex.\par}

The core of this new module consists in several functions that construct the elements of the spectral sequence of a filtered complex (groups, differential maps, and convergence levels). These main functions are:
\par
{\leftskip=0.3cm \hangafter=1 \hangindent=1.1cm \texttt{print-spct-sqn-cmpns} \textit{fltrcm r p q}\\
Display on the screen the components ($\Zset$ or $\Zset_m$) of the group $E^r_{p,q}$ of the filtered complex \textit{fltrcm}.\par}
\par
{\leftskip=0.3cm \hangafter=1 \hangindent=1.1cm \texttt{spct-sqn-basis-dvs} \textit{fltrcm r p q}\\
Return a description of the group $E^r_{p,q}$, more precisely of the numerator and denominator of the formula in Theorem \ref{thm:spct-sqn} (see details of this representation in the examples of Sections \ref{sec:examples} and \ref{sec:example2}).\par}
\par
{\leftskip=0.3cm \hangafter=1 \hangindent=1.1cm \texttt{spct-sqn-dffr} \textit{fltrcm r p q int-list}\\
Compute the differential $d^r_{p,q}:E^r_{p,q} \rightarrow E^r_{p-r,q+r-1}$ (the role of \textit{int-list} is explained in the examples).\par}
\par
{\leftskip=0.3cm \hangafter=1 \hangindent=1.1cm \texttt{spct-sqn-cnvg-level} \textit{fltrcm degr}\\
Determine the stage $r$ at which the convergence of the spectral sequence has been reached for a specific degree $degr$.\par}

To provide a better understanding of these new tools, some elementary examples of their use are showed in the next section. Besides, in Section \ref{sec:example2} we present two more interesting examples where the application of the programs allows the computation of some groups and differential maps which are beyond the calculations appearing in the literature.

These new methods work in a way that is similar to the mechanism of Kenzo for computing homology groups. If the filtered complex is effective, its spectral sequence can be determined thanks to elementary computations with the differential matrices. Otherwise, the effective homology is needed to compute it by means of the spectral sequence of the effective complex. Making use of  Theorem \ref{thm:spct-eff}, the spectral sequences of both complexes are isomorphic after some level $t$ (depending on the order of the homotopies in both reductions of the equivalence). However, we must bear in mind that in the first stages both spectral sequences are not necessarily the same.

\section{Didactic Examples}
\label{sec:examples}

As explained in the previous section, the new programs allow us to compute spectral sequences of filtered complexes with effective homology (with the exception, in some cases, of the first stages), even if the complexes are not of finite type. In this section two simple examples of this computation are presented. In these cases, the spectral sequences are well known and can be obtained without using a computer. We propose them as didactic examples for a better understanding of the new functionality.

\subsection{$S^2\times_\tau K(\Zset,1)$}

With these programs it is possible to obtain the spectral sequence of the twisted product $S^2\times_\tau K(\Zset,1)$ for a twisting operator $\tau:S^2\rightarrow K(\Zset,1)$ with  \(\tau(\texttt{s2}) = [1]\). We use here the standard simplicial description of the 2-sphere, with a unique non-degenerate simplex \texttt{s2} in dimension 2. A principal fibration is then defined by a unique 1-simplex of the simplicial structural group. The result in this case is the Hopf fibration, the total space \(S^2 \times_\tau K(\Zset, 1)\) being a simplicial model for the 3-sphere \(S^3\). The same example could be processed with \(\tau(\texttt{s2}) = [2]\), the total space then being the real projectif space \(P^3\Rset\). Let us remark that, since $K(\Zset,1)$ is not effective, the space $S^2\times_\tau K(\Zset,1)$ is not effective either, and therefore the effective homology is necessary for the computation of its spectral sequence.

To find the effective homology of the total space of a fibration $G \rightarrow B\times_\tau G \rightarrow B$, the Kenzo system needs the base and fibre spaces \(B\) and \(G\) provided with their respective effective homologies $HB$ and $HG$. The starting point is the Eilenberg-Zilber reduction $C(B\times G) \Rightarrow C(B)\otimes C(G)$ (see \cite{EZ53}). Applying the Basic Perturbation Lemma (BPL) (see \cite{RS02}) with a perturbation induced by the twisting operator $\tau$, a reduction $C(B\times_\tau G) \Rightarrow C(B)\otimes_t C(G)$ is obtained, where the symbol $\otimes_t$ represents a twisted (perturbed) tensor product, induced by $\tau$. On the other hand, from the effective homologies of $B$ and $G$,  we can construct a new equivalence from the tensorial product $C(B)\otimes C(G)$ to $HB\otimes HG$, and using again the BPL (with the perturbation to be applied to the differential of $C(B)\otimes C(G)$ to obtain the differential of $C(B)\otimes_t C(G)$) we construct an equivalence from $C(B)\otimes_t C(G)$ to a new twisted tensor product $HB\otimes_t HG$, which is an effective complex. Finally, the composition of the two equivalences is the effective homology of \(B \times_\tau G\).

Let us consider again the specific example, the twisted product $S^2\times_\tau K(\Zset,1)$. This is built in Kenzo in the following way:
\newpage
\begin{verbatim}
>(setf s2 (sphere 2))
[K23 Simplicial-Set]
>(setf kz1 (k-z 1))
[K1 Abelian-Simplicial-Group]
>(setf tau (build-smmr
           :sorc s2
           :trgt kz1
           :degr -1
           :sintr #'(lambda (dmns gmsm) (absm 0 '(1)))
           :orgn '(s2-tw-kz1)))
[K28 Fibration K23 -> K1]          
>(setf s2-tw1-kz1 (fibration-total tau))
[K34 Simplicial-Set]
\end{verbatim}

The object \texttt{tau} implements the twisting operator $\tau:S^2\rightarrow K(\Zset,1)$ as a simplicial morphism of degree $-1$ that sends the unique non-degenerate simplex \texttt{s2} of dimension $2$ to the $1$-simplex \texttt{(1)} of the simplicial set \texttt{kz1} (if we changed the list \texttt{'(1)}, that represents this $1$-simplex, by the list \texttt{'(2)}, we would obtain the Hopf fibration of the real projectif space \(P^3\Rset\)). The function \texttt{fibration-total} builds the total space of the fibration defined by the twisting operator \texttt{tau} (this operator contains as source and target spaces the base and the fibre spaces of the fibration respectively), which is a twisted cartesian product of the base and fibre.

Since the effective complex of $K(\Zset,1)$ is $S^1$, the effective complex of $S^2\times_\tau K(\Zset,1)$ will be $S^2\otimes S^1$, with an appropriate perturbation of the differential. We can inspect it by applying the function \texttt{rbcc} (\underline{r}ight \underline{b}ottom \underline{c}hain \underline{c}omplex) to the \underline{ef}fective \underline{h}o\underline{m}ology of the complex:
\begin{verbatim}
>(setf s2xts1 (rbcc (efhm s2-tw1-kz1)))
[K95 Chain-Complex]
\end{verbatim}

What is this chain complex \(K_{95}\)?
\begin{verbatim}
>(orgn s2xts1)
(ADD [K74 Chain-Complex] [K93 Morphism (degree -1): K74 -> K74])
\end{verbatim}

This origin means that the complex \texttt{s2xts1} has been obtained by application of the BPL, ``adding'' a perturbation (the morphism \(K_{93}\), of degree $-1$ ) to the initial chain complex \(K_{74}\). We want to know now what \(K_{74}\) is:
\begin{verbatim}
>(orgn (k 74)) 
(TNSR-PRDC [K23 Simplicial-Set] [K16 Chain-Complex])
\end{verbatim}

As expected, we have a \underline{t}e\underline{ns}o\underline{r} \underline{pr}o\underline{d}u\underline{c}t. And finally, what about \(K_{23}\) and \(K_{16}\)?
\begin{verbatim}
> (orgn (k 23))
(SPHERE 2)
> (orgn (k 16))
(CIRCLE)
\end{verbatim}

In this way we can state that \(K_{23}=S^2\) and \(K_{16}=S^1\), and therefore the effective complex of \texttt{s2-tw1-kz1}$=S^2\times_\tau K(\Zset,1)$ is $S^2\otimes S^1$ with a perturbation of the differential.

To compute the spectral sequence of this twisted product (Serre spectral sequence, \cite{SER51}), it is necessary to change it into a filtered complex. The filtration in this complex is defined through the degeneracy degree with respect to the base space: a generator $(x_n,y_n)\in C(B\times G)$ has a filtration degree less or equal to $q$ if $\exists \bar{y}_q\in B_q$ such that $y_n=s_{i_{n-q}}\cdots s_{i_1}\bar{y}_q$. Such a filtration can be implemented as follows.
\begin{verbatim}
>(setf twpr-flin 
      #'(lambda (degr crpr)
          (declare
           (type fixnum degr)
           (type crpr crpr))
          (let* ((b (cadr crpr))
                 (dgop (car b)))
             (declare
              (type iabsm b)
              (type fixnum dgop))
             (the fixnum
               (- degr (length (dgop-int-ext dgop)))))))
>(CHANGE-chcm-TO-FltrChcm s2-tw1-kz1 :flin twpr-flin 
    :orgn `(filtered-complex ,s2-tw1-kz1))
[K34 Filtered-Complex]    
\end{verbatim} 

A filtration is also needed in the effective complex, $S^2\otimes_t S^1$, which is filtered by the base dimension. In general, for a tensor product: $F_p(C(B)\otimes C(G))=\oplus_{m\leq p}C(B)_m \otimes C(F)$. The implementation in Common Lisp is as follows.
\begin{verbatim}
>(setf tnpr-flin 
      #'(lambda (degr tnpr)
       (declare 
         (type fixnum degr)
         (type tnpr tnpr))
       (the fixnum
          (degr1 tnpr))))
\end{verbatim}
\newpage
\begin{verbatim}
> (CHANGE-chcm-TO-FltrChcm s2xts1 :flin tnpr-flin 
    :orgn `(filtered-complex ,s2xts1))
[K95 Filtered-Complex]
\end{verbatim} 

Once the filtrations are defined, the new programs can be used to compute the spectral sequence of the twisted product $S^2\times_\tau K(\Zset,1)$, which is isomorphic in every level to that of the effective complex $S^2\otimes_t S^1$ because in this case both homotopies in the equivalence have order equal to zero. For instance, the groups $E^2_{2,0}$ and $E^2_{0,1}$ are equal to $\Zset$: 
\begin{verbatim}
> (print-spct-sqn-cmpns s2-tw1-kz1 2 2 0)
Spectral sequence E^2_{2,0}
Component Z
> (print-spct-sqn-cmpns s2-tw1-kz1 2 0 1)
Spectral sequence E^2_{0,1}
Component Z
\end{verbatim}

These groups can be recognized as the elements of the Serre spectral sequence of the Hopf fibration.

It is also possible to find the \textit{basis-divisors} representation of these groups. This representation shows a list of combinations which generate the subgroup in the \emph{numerator} of $E^r_{p,q}$ ($Z^r_{p,q} \cup F_{p-1}C_{p+q}$), as well as the coefficients (with regard to this list of combinations) of the elements that generate the \emph{denominator} ($dZ^{r-1}_{p+r-1,q-r+2}\cup F_{p-1}C_{p+q}$). For the groups $E^2_{2,0}$ and $E^2_{0,1}$ that have been computed above:
\begin{verbatim}
>(spct-sqn-basis-dvs s2-tw1-kz1 2 2 0)
((
  ------------------------------------------------------{CMBN 2}
  <-1 *   <CrPr - S2 1-0 NIL>>
  --------------------------------------------------------------
    )
   (0))
> (spct-sqn-basis-dvs s2-tw1-kz1 2 0 1)
((
  ------------------------------------------------------{CMBN 1}
  <-1 *   <CrPr 0 * - (1)>>
  --------------------------------------------------------------
    )
   (0))
\end{verbatim}

In both cases, the ``basis'' (list of combinations) of the numerator has a unique element and the list of ``divisors'' is the list \texttt{(0)}. This means that the subgroup of the numerator is isomorphic to $\Zset$ and that of the denominator is the null group $0$, so that the wanted $E^r_{p,q}$ is in both cases isomorphic to $\Zset$. For $E^2_{2,0}$, the generator is the element \(-1*(\texttt{s2}, \eta_1 \eta_0 [\ ]) \in S^2 \times_\tau K(\Zset, 1)\), which is not a torsion element. In a similar way, the unique generator of $E^2_{0,1}$ is \(-1*(\eta_0 \ast, [1])\), not a torsion element either. If the second component of a result were, for instance, \texttt{(3)} instead of \texttt{(0)}, the denominator generator would be 3 times the numerator generator and the corresponding \(E^r_{p,q}\) would be the torsion group \(\Zset_3\).    

The differential function in a group $E^r_{p,q}$ can be computed using the function \texttt{spct-sqn-dffr}. The last argument must be a list that represents the coordinates of the element we want to apply the differential to (with regard to the generators of the subgroup in the numerator). In the example that follows, the differential $d^2_{2,0}$ is applied to the generator of the group $E^2_{2,0}\cong \Zset$ (that, as we have seen, is the following combination of degree $2$: \(-1*(\texttt{s2}, \eta_1 \eta_0 [\ ])\)), and therefore the list of coordinates must be \texttt{(1)} (the list \texttt{(2)}, for instance, would correspond to the combination \(-2*(\texttt{s2}, \eta_1 \eta_0 [\ ])\)).
\begin{verbatim}
> (spct-sqn-dffr s2-tw1-kz1 2 2 0 '(1))
(1)
\end{verbatim}
The obtained list \texttt{(1)} shows that the result of applying $d^2_{2,0}$ to the generator of the group is the combination $1 * g^2_{0,1}$, where $g^2_{0,1}$ is the generator of the group $E^2_{0,1}\cong\Zset$ (which is the combination of degree $1$: \(-1*(\eta_0 \ast, [1])\)). This last result means that the differential map $d^2_{2,0}: E^2_{2,0} \rightarrow E^2_{0,1}$ maps \((\texttt{s2}, \eta_1 \eta_0 [\ ])\) to \((\eta_0 \ast, [1])\). Since the next stage in the spectral sequence $E^3$ is isomorphic to the bigraded homology group of $E^2$, $E^3_{p,q}\cong H_{p,q}(E^2)=\mathrm{Ker}\ d^2_{p,q} / \  \mathrm{Im}\ d^2_{p+2,q-1}$, it is clear that the groups $E^3_{0,1}$ and $E^3_{2,0}$ must be null:
\begin{verbatim}
> (print-spct-sqn-cmpns s2-tw1-kz1 3 0 1)
Spectral sequence E^3_{0,1}
NIL
> (print-spct-sqn-cmpns s2-tw1-kz1 3 2 0)
Spectral sequence E^3_{2,0}
NIL
\end{verbatim}

Finally, it is also possible to obtain, for each degree $n$, the level $r$ at which the convergence of the spectral sequence has been reached, that is, the smallest $r$ such that $E^\infty_{p,q}=E^r_{p,q} \ \forall p,q$ with $p+q=n$. For instance, for $n=0$ and $n=1$ the convergence levels are $1$ and $3$ respectively:
\begin{verbatim}
>(spct-sqn-cnvg-level s2-tw1-kz1 0)
1
>(spct-sqn-cnvg-level s2-tw1-kz1 1)
3
\end{verbatim}

Thus, we can obtain the groups $E^\infty_{p,q}$ with $p+q=0$ or $p+q=1$ by computing the corresponding groups $E^1_{0,0}$, $E^3_{0,1}$, and $E^3_{1,0}$:
\begin{verbatim}
> (print-spct-sqn-cmpns s2-tw1-kz1 1 0 0)
Spectral sequence E^1_{0,0}
Component Z
> (print-spct-sqn-cmpns s2-tw1-kz1 3 0 1)
Spectral sequence E^3_{0,1}
NIL
> (print-spct-sqn-cmpns s2-tw1-kz1 3 1 0)
Spectral sequence E^3_{1,0}
NIL
\end{verbatim}
  
\subsection{$S^2\times_\tau K(\Zset_2,1)$}

Another example, similar to the previous one, is the twisted product $S^2\times_\tau K(\Zset_2,1)$, with $\tau:S^2\rightarrow K(\Zset_2,1)$, $\tau(s2)=[1]$. Let us consider our new structural group, the simplicial group $K(\Zset_2,1)$: in dimension $n$, the only non-degenerate simplex is a sequence of $n$ 1's, represented by the integer $n$, and the integer $0$ encodes the void bar object $[ ]$. The twisted product is implemented in the same way as the first example, using the function \texttt{fibration-total} to build the total space of the fibration defined by the morphism \texttt{tau2}. Afterwards the complex is provided with the usual filtration for twisted products (through the degeneracy degree with respect to the base space), implemented in the function \texttt{twpr-flin} of the previous subsection.  
\begin{verbatim}
>(setf kz21 (k-z2 1))
[K110 Abelian-Simplicial-Group]
>(setf tau2 (build-smmr
           :sorc s2
           :trgt kz21
           :degr -1
           :sintr #'(lambda (dmns gmsm) (absm 0 1))
           :orgn '(s2-tw-kz2)))
[K122 Fibration K23 -> K110]
>(setf s2-tw2-kz21 (fibration-total tau2))
[K128 Simplicial-Set]
>(CHANGE-chcm-TO-FltrChcm s2-tw2-kz21 :flin twpr-flin 
    :orgn `(filtered-complex ,s2-tw2-kz21))
[K128 Filtered-Complex]    
\end{verbatim}

In this case the complex is finitely generated (in each degree), and therefore the spectral sequence can be computed directly without the need of effective homology. For instance, some groups of the spectral sequence are:
\newpage
\begin{verbatim}
> (print-spct-sqn-cmpns s2-tw2-kz21 2 0 1)
Spectral sequence E^2_{0,1}
Component Z/2Z
> (print-spct-sqn-cmpns s2-tw2-kz21 2 2 0)
Spectral sequence E^2_{2,0}
Component Z
> (print-spct-sqn-cmpns s2-tw2-kz21 3 0 3)
Spectral sequence E^3_{0,3}
Component Z/2Z
\end{verbatim}

The groups $E^2_{0,1}$ and $E^3_{0,3}$ are isomorphic to $\Zset_2$ and therefore their generators must be torsion elements, with order $2$. We can inspect, for instance, the generator of $E^2_{0,1}$:
\begin{verbatim}
>(spct-sqn-basis-dvs s2-tw2-kz21 2 0 1) 
((
  ------------------------------------------------------{CMBN 1}
  <1 *   <CrPr 0 * - 1>>
  --------------------------------------------------------------
    )
   (2))
\end{verbatim}
The combination \(1*(\eta_0 \ast, 1)=(\eta_0 \ast, 1)\) is the generator of the numerator in $E^2_{0,1}$. The list of divisors \texttt{(2)} means that the denominator in the quotient is generated by the element $2*g$, where $g$ is the respective generator in the numerator, in this case \(g=(\eta_0 \ast, 1)\). In this way, the group $E^2_{0,1}=\Zset(\eta_0 \ast, 1)/\Zset(2*(\eta_0 \ast, 1))\cong \Zset/ 2\Zset\cong\Zset_2$. The generator of the group $E^2_{0,1}$ is therefore the element \((\eta_0 \ast, 1)\), with order $2$.

Similarly, we can obtain the generator of the group $E^2_{2,0}\cong\Zset$:
\begin{verbatim}
>(spct-sqn-basis-dvs s2-tw2-kz21 2 2 0)
((
  ------------------------------------------------------{CMBN 2}
  <1 *   <CrPr - S2 1 1>>
  <-1 *   <CrPr - S2 1-0 0>>
  <1 *   <CrPr 1-0 * - 2>>
  --------------------------------------------------------------
     
  ------------------------------------------------------{CMBN 2}
  <1 *     <CrPr - S2 - 2>>
  <-1 *     <CrPr - S2 0 1>>
  --------------------------------------------------------------
\end{verbatim}
\newpage
\begin{verbatim}  
  ------------------------------------------------------{CMBN 2}
  <-1 *     <CrPr - S2 0 1>>
  <1 *     <CrPr - S2 1-0 0>>
  --------------------------------------------------------------
     
  ------------------------------------------------------{CMBN 2}
  <-1 *     <CrPr - S2 1 1>>
  <1 *     <CrPr - S2 1-0 0>>
  --------------------------------------------------------------
     
  ------------------------------------------------------{CMBN 2}
  <1 *     <CrPr - S2 1-0 0>>
  --------------------------------------------------------------
    )
   (1 1 1 1 0))
\end{verbatim}

In this case the numerator is generated by five elements, the combinations $1*(\texttt{s2},\eta_1 1)-1*(\texttt{s2},\eta_1\eta_0 0)+1*(\eta_1\eta_0\ast,2),1*(\texttt{s2},2)-1*(\texttt{s2},\eta_0 1),-1*(\texttt{s2},\eta_0 1)+1*(\texttt{s2},\eta_1\eta_0 0),-1*(\texttt{s2},\eta_1 1)+1*(\texttt{s2},\eta_1\eta_0 0)$, and $1*(\texttt{s2},\eta_1\eta_0 0)$. The denominator has four generators, the four first combinations in the previous list: $1*(\texttt{s2},\eta_1 1)-1*(\texttt{s2},\eta_1\eta_0 0)+1*(\eta_1\eta_0\ast,2),1*(\texttt{s2},2) -1*(\texttt{s2},\eta_0 1),-1*(\texttt{s2},\eta_0 1)+1*(\texttt{s2},\eta_1\eta_0 0)$, and $-1*(\texttt{s2},\eta_1 1)+1*(\texttt{s2},\eta_1\eta_0 0)$. Therefore the group $E^2_{2,0}$, that is isomorphic to $\Zset$, is generated by the fifth combination, the element \(1*(\texttt{s2},\eta_1\eta_0 0)=(\texttt{s2},\eta_1\eta_0 0)\).
 
As in the previous example, it is possible to compute the differential of this group applied, for instance, to the generator \((\texttt{s2},\eta_1\eta_0 0)\) (whose coordinates correspond to the list \texttt{(1)}):
\begin{verbatim}
> (spct-sqn-dffr s2-tw2-kz21 2 2 0 '(1))
(1)
\end{verbatim}
This means that $d^2_{2,0}((\texttt{s2},\eta_1\eta_0 0))=(\eta_0 \ast, 1)$.

Finally, some convergence levels (for $n=1,2,3$) are:
\begin{verbatim}
> (spct-sqn-cnvg-level s2-tw2-kz2 1)
3
> (spct-sqn-cnvg-level s2-tw2-kz2 2)
1
> (spct-sqn-cnvg-level s2-tw2-kz2 3)
1
\end{verbatim}

\section{Advanced examples}
\label{sec:example2}

As said in the first paragraph of Section \ref{sec:examples}, the two filtered complexes presented there are elementary and the computation of their spectral sequences can be done by hand without any special difficulty. We introduce in this section two other examples, which are perhaps not so easy to understand as the preceding ones, but they have a higher interest because their spectral sequences seem difficult to be studied by the theoretical methods documented in the literature. However, with the use of the new programs the different groups $E_{p,q}^r$ and the differential maps $d^r_{p,q}$ are computed.

\subsection{Postnikov tower}

The first example considered in this section corresponds to the space $X_4$ of a Postnikov tower \cite{May67} with a $\pi_i=\Zset_2$ at each stage and the ``simplest'' non-trivial Postnikov invariant. Ours programs compute the groups $E^r_{p,q}$ of the Serre spectral sequence of the fibration producing our space \(X_4\) in a short time for $p+q<6$, and they determine some differential maps $d^5$ which are not null.

The theoretical details of the construction of the space $X_4$ are not included here, they can be found in \cite[pp. 142-145]{RS05}. This complex can be built by Kenzo with the following statements:

\begin{verbatim}
> (setf X2 (k-z2 2))
[K133 Abelian-Simplicial-Group]
> (setf k3 (chml-clss X2 4))
[K245 Cohomology-Class on K150 of degree 4]
> (setf F3 (z2-whitehead X2 k3))
[K260 Fibration K133 -> K246]
> (setf X3 (fibration-total F3))
[K266 Kan-Simplicial-Set]
> (setf k4 (chml-clss X3 5))
[K479 Cohomology-Class on K464 of degree 5]
> (setf F4 (z2-whitehead X3 k4))
[K494 Fibration K266 -> K480]
> (setf X4 (fibration-total F4))
[K500 Kan-Simplicial-Set]
\end{verbatim}

This example also corresponds to a total space of a fibration, a twisted product $K(\Zset_2,4) \times_{k_4} X_3$, where $X_3$ is again a twisted product $K(\Zset,3) \times_{k_3} K(\Zset_2,2) $ and $k_4$ and $k_3$ are called the \emph{k-invariants} of the Postnikov tower. Therefore the filtrations in the space $X_4$ and in its effective complex are defined as in the examples of Section \ref{sec:examples} (since all of them are particular instances of the Serre spectral sequence):
\begin{verbatim}
> (setf effX4 (rbcc (efhm X4)))
[K696 Chain-Complex]
> (CHANGE-CHCM-TO-FltrChcm X4 :flin fbrt-flin 
     :orgn `(filtered-complex ,X4))
[K500 Filtered-Complex]
> (CHANGE-CHCM-TO-FltrChcm effX4 :flin tnpr-flin 
     :orgn `(filtered-complex ,effX4))
[K696 Filtered-Complex]     
\end{verbatim}

Some groups $E^r_{p,q}$ at the stage $r=2$ are:

\begin{verbatim}
>(print-spct-sqn-cmpns X4 2 0 4)
Spectral sequence E^2_{0,4}
Component Z/2Z
> (print-spct-sqn-cmpns X4 2 5 0)
Spectral sequence E^2_{5,0}
Component Z/4Z
> (print-spct-sqn-cmpns X4 2 6 0)
Spectral sequence E^2_{6,0}
Component Z/2Z
Component Z/2Z
\end{verbatim}

For $p+q=4,5,6,7$, the spectral sequence converges at the stage $r=6$:
\begin{verbatim}
> (spct-sqn-cnvg-level X4 4)
6
> (spct-sqn-cnvg-level X4 5)
6
> (spct-sqn-cnvg-level X4 6)
6
> (spct-sqn-cnvg-level X4 7)
6
\end{verbatim}

This means that there are some differential maps $d^5$ which are not null. Specifically, the programs compute $d^5_{5,0}$ and $d^5_{7,0}$ that map the unique generators of $E^5_{5,0}\cong\Zset_4$ and $E^5_{7,0}\cong\Zset_2$ to the unique generators of $E^5_{0,4}$ and $E^5_{2,4}$ (both isomorphic to $\Zset_2$) respectively:
\begin{verbatim}
>  (spct-sqn-dffr X4 5 5 0 '(1))
(1)
>  (spct-sqn-dffr X4 5 7 0 '(1))
(1)
\end{verbatim}

Finally, we can conclude that the groups $E^\infty_{p,q}$ for $p+q=4,5,6,7$ are the same than $E^6_{p,q}$, which are easily obtained. For instance, for $n=7$, all of them are null except $E^6_{0,7}$ and $E^6_{3,4}$):
\begin{verbatim}
> (dotimes (p 8)
         (let ((q (- 7 p)))
            (terpri)
            (print-spct-sqn-cmpns X4 6 p q)
            ))
Spectral sequence E^6_{0,7}
Component Z/2Z
Spectral sequence E^6_{1,6}
Spectral sequence E^6_{2,5}
Spectral sequence E^6_{3,4}
Component Z/2Z
Spectral sequence E^6_{4,3}
Spectral sequence E^6_{5,2}
Spectral sequence E^6_{6,1}
Spectral sequence E^6_{7,0}
\end{verbatim}

\subsection{Eilenberg-Moore spectral sequence}

The programs presented here can also be used to determine the Eilenberg-Moore spectral sequence between a simplicial set $X$ and its loop space $\Omega X$, whose theoretical definition can be found in \cite{EM66}. If the space $X$ is an $1$-reduced simplicial set with effective homology, the program Kenzo determines the effective homology of its loop space $\Omega X$ using the cobar construction on a coalgebra. Moreover, if $X$ is $m$-reduced, this process may be iterated $m$ times, producing an effective homology version of $\Omega^kX, k\leq m$. The effective homology of the loop space, together with the natural filtration defined on the cobar construction allows the computation by our programs of the spectral sequence between $H_*(X)$ and $H_*(\Omega X)$.

The Eilenberg-Moore spectral sequence has been traditionally considered to be an important tool for obtaining homotopic information of a space, by means of its relation with its loop space. In particular, it can be used for the study of the effect of the attachment of a disk to an space of infinite dimension, especially a loop space. This problem seems to be very difficult in general as explained in \cite{Lem78}. Our programs have determined the different elements of the spectral sequence for some spaces constructed in this way that, up to now, have not appeared in the literature. As a little introduction of this work (that is yet incomplete and whose details will appear in a future paper), we present in Figures 1 and 2 the groups $E^\infty_{p,q}$ (for $q-p\leq 8$) of the spectral sequences for the spaces $\Omega S^3$ and $\Omega S^3 \cup_2 D^3$ (the last one obtained from $\Omega S^3$ by attaching a $3$-disk by a map \(\gamma: S^2 \rightarrow \Omega S^3\) of degree 2). The first space and its loop space have been extensively considered by theoretical methods and a lot of results about them are known. However, for our second example, the attachment of the $3$-disk increases the difficulty of the calculation of the Eilenberg-Moore spectral sequence between $\Omega S^3 \cup_2 D^3$ and its loop space that, up to our knowledge, had not been determined before. See Figures 1 and 2 for the calculated \(E^r_{p,q}\)'s.

\begin{center}
\begin{figure}
\label{fig1}
\caption{Groups $E^\infty_{p,q}$ of the Eilenberg-Moore spectral sequence between $\Omega S^3$ and $\Omega(\Omega S^3)$}
\frame{

\xymatrix @R=0.6mm @C=0.5cm {
& & \\
 &  & \ar @{-}[ddddddddddddddddddddd] \\
\\& q
\\
\\
& 16 & &   &   &   &   &   &   &   &   & \Zset_2  &\\ 
& 15 & &   &   &   &   &   &   &  & 0\\
& 14 & &   &   &   &   &   &   & 0 & \Zset_2  &\\
& 13 & &   &   &   &   &   & 0 & 0\\
& 12 & &   &   &   &   & \Zset_6 & 0 & \Zset_2  &\\
& 11 & &   &   &   & 0 & 0 & 0\\
& 10 & &  &  & \Zset_5 & \Zset_2\ & 0 & \Zset_2\\
& 9 & &   & 0 & 0 & 0 & 0 \\
& 8 & & 0 & 0 & \Zset_2 & \Zset_3 & \Zset_2\\
& 7 & & 0 & 0 & 0 & 0\\
& 6 & & 0 & 0 & \Zset_3 & \Zset_2 \\
& 5 & & 0 & 0 & 0\\
& 4 & & 0 & 0 & \Zset_2 \\
& 3 & & 0 & 0 \\
& 2 & & 0 & \Zset\\
& 1 & & 0  \\
& 0 & & \Zset \\
& &  \ar @{-}[rrrrrrrrrrrr] & & & & & & & & & & & & &\\
& & & 0 & 1 & 2 & 3 & 4 & 5 & 6 & 7 & 8 & & p  &\\
& & &
}}

\end{figure}
\end{center}

\begin{center}
\begin{figure}
\label{fig2}
\caption{Groups $E^\infty_{p,q}$ of the Eilenberg-Moore spectral sequence between $\Omega S^3\cup_2 D^3$ and $\Omega(\Omega S^3\cup_2 D^3)$}
\frame{

\xymatrix @R=0.6mm @C=0.4cm {
& & \\
 &  & \ar @{-}[ddddddddddddddddddddd] \\
\\& q
\\
\\
& 16 & &   &   &   &   &   &   &   &   & \Zset_2  &\\ 
& 15 & &   &   &   &   &   &   &   & \Zset_2^6\\
& 14 & &   &   &   &   &   &   & \Zset_2^{11} & \Zset_2  &\\
& 13 & &   &   &   &   &   & \Zset_2^{11} & \Zset_5\\
& 12 & &   &   &   &   & \Zset_2^5 & \Zset_2^7 & \Zset_2  &\\
& 11 & &   &   &   & 0 & \Zset_2^6 & \Zset_2^4\\
& 10 & &    &   & \Zset_2\oplus\Zset_{10}\oplus\Zset & \Zset_2\ & \Zset_2^4 & \Zset_2\\
& 9 & &   & 0 & 0 & \Zset_2^3 & \Zset_2^3 \\
& 8 & & 0 & 0 & \Zset_6 & \Zset_2^2 & \Zset_2\\
& 7 & & 0 & 0 & 0 & \Zset_2^2\\
& 6 & & 0 & \Zset & \Zset_2 & \Zset_2 \\
& 5 & & 0 & 0 & \Zset_2\\
& 4 & & 0 & \Zset & \Zset_2 \\
& 3 & & 0 & 0 \\
& 2 & & 0 & \Zset_2\\
& 1 & & 0  \\
& 0 & & \Zset \\
& &  \ar @{-}[rrrrrrrrrrrr] & & & & & & & & & & & & \\
& & & 0 & 1 & 2 & 3 & 4 & 5 & 6 & 7 & 8 & & p  &\\
& & &
}}

\end{figure}
\end{center}

\section{Conclusions and further work}

In this note, we have presented some programs that improve the functionality of Kenzo, computing spectral sequences (groups, differential maps, and convergence levels) of filtered complexes with effective homology. These programs can be applied to compute, for instance, spectral sequences of double complexes, the Serre spectral sequence, the Eilenberg-Moore spectral sequences...

One of our next goals is the development of some new programs (working in a similar way to that explained in this paper) to deal with exact couples \cite{WHI78}. As it is known, exact couples determine spectral sequences (containing more information that makes it possible to determine the successive differentials $d^r$). This is a way of obtaining spectral sequences more general than that of filtered complexes: a filtered complex determines an exact couple whose spectral sequence is isomorphic to that of the filtered complex; on the other hand, an exact couple does not need to arise from a filtration. In this way, there are spectral sequences (for instance, the Bousfield-Kan spectral sequence) which do not correspond to any filtered complex. Therefore we find that it would be interesting to build a new set of programs allowing its computation.

\end{document}